\documentclass[floatfix, showkeys, reprint, superscriptaddress, amsmath, amssymb, aps]{revtex4-1}

\usepackage[utf8]{inputenc}
\usepackage[T1]{fontenc}
\usepackage{graphicx}
\usepackage{dcolumn}
\usepackage{bm}
\usepackage{braket}
\usepackage{caption}
\usepackage[caption=false]{subfig}
\usepackage{float}
\usepackage{booktabs}
\usepackage{multirow}
\usepackage[normalem]{ulem}
\usepackage{soul}
\usepackage{xcolor}
\usepackage{ragged2e}

\usepackage{comment}

\begin{document}

\preprint{APS/123-QED}

\title{Molecular Docking with Quantum Circuit Evolution}

\author{Gleydson F. de Jesus}
\email{gleydson.jesus@fieb.org.br}
\affiliation{QuIIN—Quantum Industrial Innovation, EMBRAPII CIMATEC Competence Center in Quantum Technologies, SENAI CIMATEC, Av. Orlando Gomes, 1845, Salvador, Bahia, 41850-010, Brazil}
\affiliation{Latin America Quantum Computing Center, SENAI CIMATEC, Salvador 41650-010, Bahia, Brazil}
\affiliation{Computational Modeling and Industrial Technology Program, SENAI CIMATEC University, Salvador, Bahia 41650-010, Brazil}

\author{Bruno O. Fernandez}
\noaffiliation

\author{Marcelo A. Moret}
\email{moret@fieb.org.br}
\affiliation{QuIIN—Quantum Industrial Innovation, EMBRAPII CIMATEC Competence Center in Quantum Technologies, SENAI CIMATEC, Av. Orlando Gomes, 1845, Salvador, Bahia, 41850-010, Brazil}
\affiliation{Computational Modeling and Industrial Technology Program, SENAI CIMATEC University, Salvador, Bahia 41650-010, Brazil}

\date{\today}

\begin{abstract}

Molecular docking is an important step in drug discovery, enabling the evaluation of receptor-ligand affinity while reducing experimental costs and increasing the number of possible tests. However, the high computational cost associated with molecular docking remains a limiting factor that can restrict both the experimental precision and the scale of the problems being addressed. To improve the future applicability of molecular docking, recent works have proposed the use of quantum algorithms based on Gaussian Boson Sampling quantum computers and also gate-based quantum computers. In this work, we propose the use of Quantum Circuit Evolution (QCE) for solving the molecular docking problem, a gate based and gradient-free quantum evolutionary method whose evolution is driven by the random application of unitary operations to a quantum circuit. The proposed algorithm demonstrated the ability to find the best solution to the problem in fewer steps than the methods presented in previous studies, exhibiting fast and stable convergence.

\end{abstract}

\maketitle

\section{Introduction}\label{sec1}
\label{introduction}

Protein–ligand interactions lie at the core of understanding drug action \cite{ijms17020144}. Through the interaction between ligands and receptors, it is possible to modulate molecular activities, thereby producing therapeutic effects against diseases or even contributing to their onset and progression \cite{Jin2025Nuclear, receptor-ligand-JCR, editorial-protein-ligand}.

In this context, the computational simulation of these interactions is one of the most important steps in drug design and discovery. It enables the preliminary screening of therapeutic molecules, increasing the number of candidates that can be evaluated while reducing the financial costs associated with drug testing and development process \cite{docking-review1}.

On the other hand, the computational cost of simulating protein–ligand interactions is high, and limited computational resources may compromise the quality of the simulations. For this reason, computational heuristics such as genetic algorithms \cite{AutodDock-Vina}, simulated annealing \cite{simulated-annealing}, and deep neural networks \cite{McNutt2021-nx} have been employed, as they can reduce the computational cost of simulations and facilitate the search for binding poses.

More recently, quantum computing approaches have been proposed, including methods based on Gaussian Boson Sampling quantum computers \cite{doi:10.1126/sciadv.aax1950, Yu2023-kc} and gate-based quantum computers \cite{PhysRevApplied.21.034036}. However, the proposed methods still face challenges associated with the classical optimization stages of quantum algorithms.

In this manuscript, we propose the use of Quantum Circuit Evolution (QCE) \cite{franken2020, Fernandez2026}, a gradient-free evolutionary method, for molecular docking simulations. This method is inspired by classical genetic algorithms and differs from previously proposed quantum approaches in that it does not require the classical optimization of quantum circuit parameters.

To enable a direct comparison between the proposed method and previously reported gate-based approaches, we adopted the problem formulation presented in \cite{PhysRevApplied.21.034036}, which considers systems of 6, 8, and 12 qubits using samples extracted from the PDBbind database \cite{Wang2004-vo, Wang2005-gn, Z.Liu2015}.

For the 6-qubit case, the problem was constructed based on the interaction between SARS-CoV-2 Mpro and PM-2-020B (PDB ID: 8SKH). For the 8-qubit case, the formulation was based on the interaction between DPP-4 and piperidine-fused imidazopyridine 34 (PDB ID: 3HAC). Finally, for the 12-qubit case, the problem was constructed from the interaction between HIV-1 gp120 and JP-III-048 (PDB ID: 5F4L).

\section{Methods}
\label{methods}

\subsection{Molecular Docking}
\label{docking}

The approach used in this manuscript to address the molecular docking problem is based on the model proposed in \cite{doi:10.1126/sciadv.aax1950, PhysRevApplied.21.034036}. The problem formulation begins with the selection of pharmacophores, atomic groups that are more likely to participate in molecular interactions. In this manuscript, as in previous works, six pharmacophore types are considered: hydrogen-bond donors/acceptors, aromatic rings, hydrophobic centroids, cations, and anions.

The identification of these atomic groups can be facilitated through the use of heuristics and software packages such as RDKit \cite{rdkit}.

After selecting the pharmacophores in the receptor and ligand, Labeled Distance Graphs (LDGs) are constructed. In these graphs, vertices represent the selected pharmacophores and are labeled according to their type, while edges are weighted by the Euclidean distances between them. The interaction between the receptor graph ($LDG_r$) and the ligand graph ($LDG_l$) then gives rise to a Bipartite Interaction Graph (BIG), whose edges encode potential interactions between receptor and ligand pharmacophores, thereby describing a predicted protein–ligand binding profile.

In this way, the original problem of identifying biological interactions is transformed into a Maximum Vertex Weight Clique Problem (MVWCP), which can be addressed computationally.

After constructing the LDGs, the compatibility between interactions must be evaluated based on the distances between the atomic sites in $LDG_1$ and $LDG_2$, according to Eq. (\ref{eq:distancias}).

\begin{equation}\label{eq:distancias}
  |d_1 - d_2| \geq \tau + \epsilon_1 + \epsilon_2
\end{equation}

where $d_1$ and $d_2$ are the corresponding distances in $LDG_1$ and $LDG_2$, respectively, and $\tau$, $\epsilon_1$, and $\epsilon_2$ are parameters that define a flexibility margin for binding affinity.

A binding interaction is considered feasible if it satisfies Eq. (\ref{eq:distancias}). When $|d_1 - d_2| = \tau + \epsilon_1 + \epsilon_2$, a perfect match is obtained. When $|d_1 - d_2| < \tau + \epsilon_1 + \epsilon_2$, a $\tau$-flexible match occurs. Conversely, when $|d_1 - d_2| > \tau + \epsilon_1 + \epsilon_2$, the interaction is considered incompatible.

\begin{figure}[H]
  \centering
  \subfloat[Match: $d_1 = d_2$.\label{fig:match}]{%
    \includegraphics[width=0.46\linewidth]{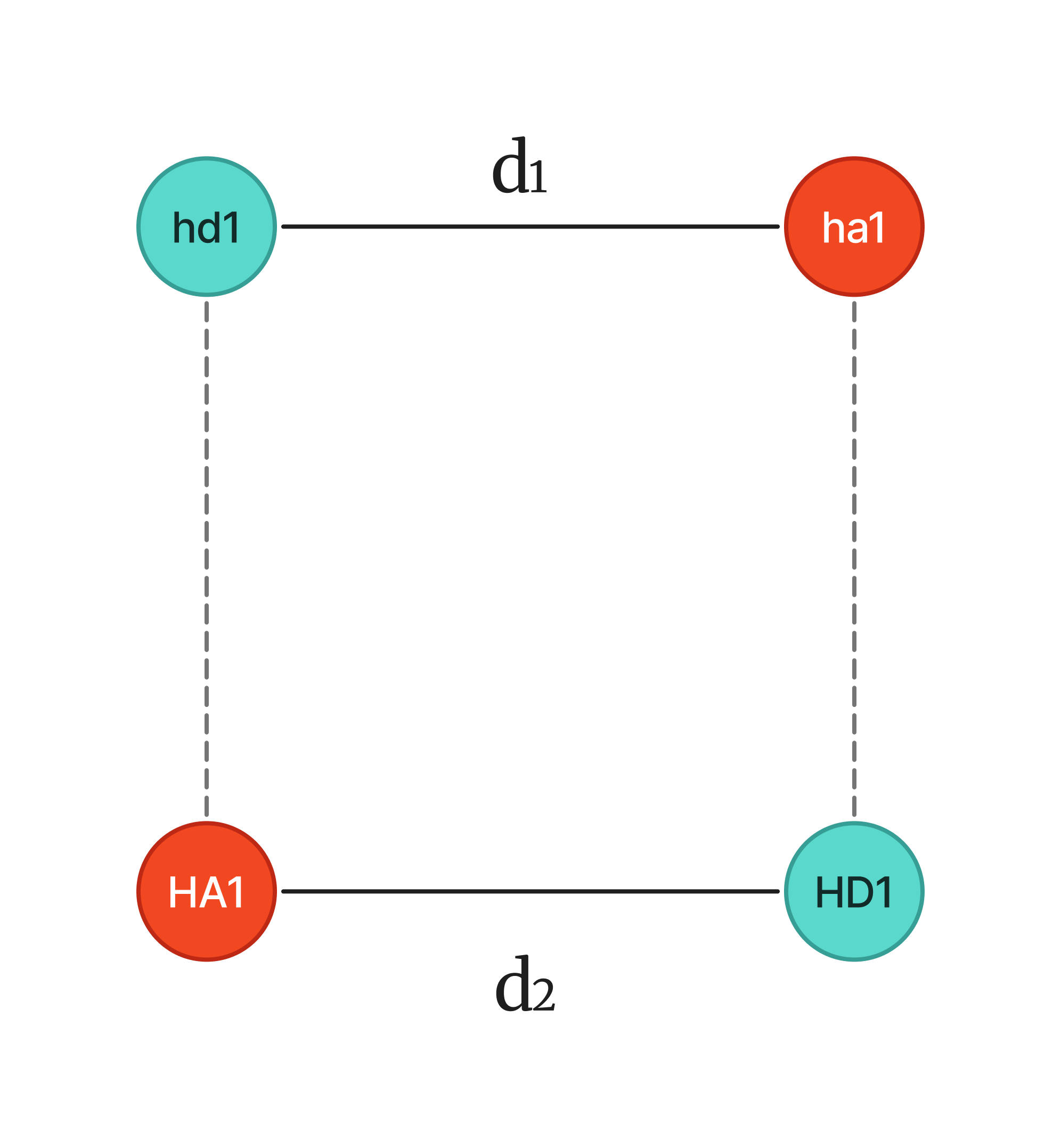}}
  \hfill
  \subfloat[$\tau$-flexible match.\label{fig:flexible-match}]{%
    \includegraphics[width=0.50\linewidth]{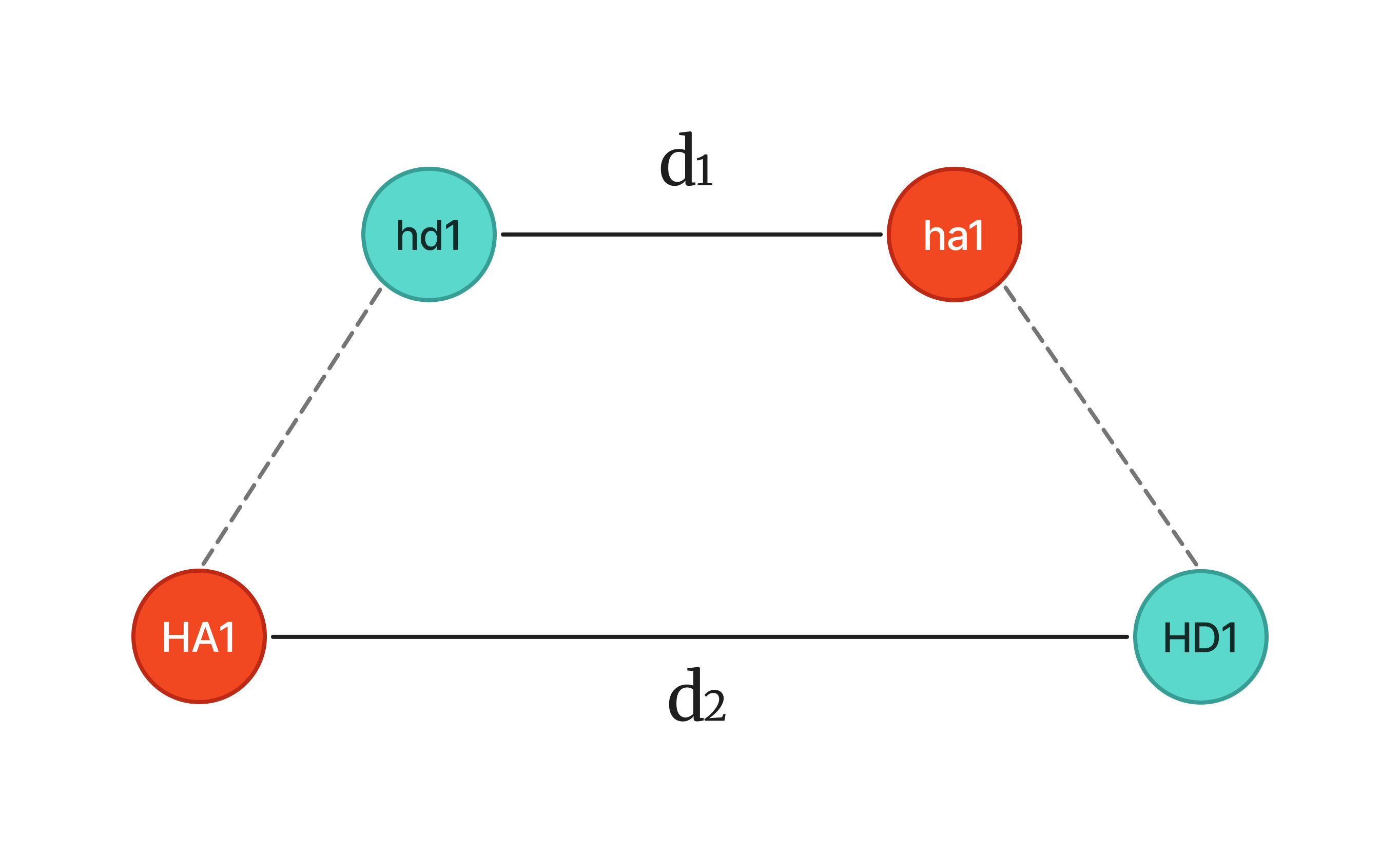}}
  \par\smallskip
  \subfloat[Mismatch.\label{fig:mismatch}]{%
    \includegraphics[width=0.62\linewidth]{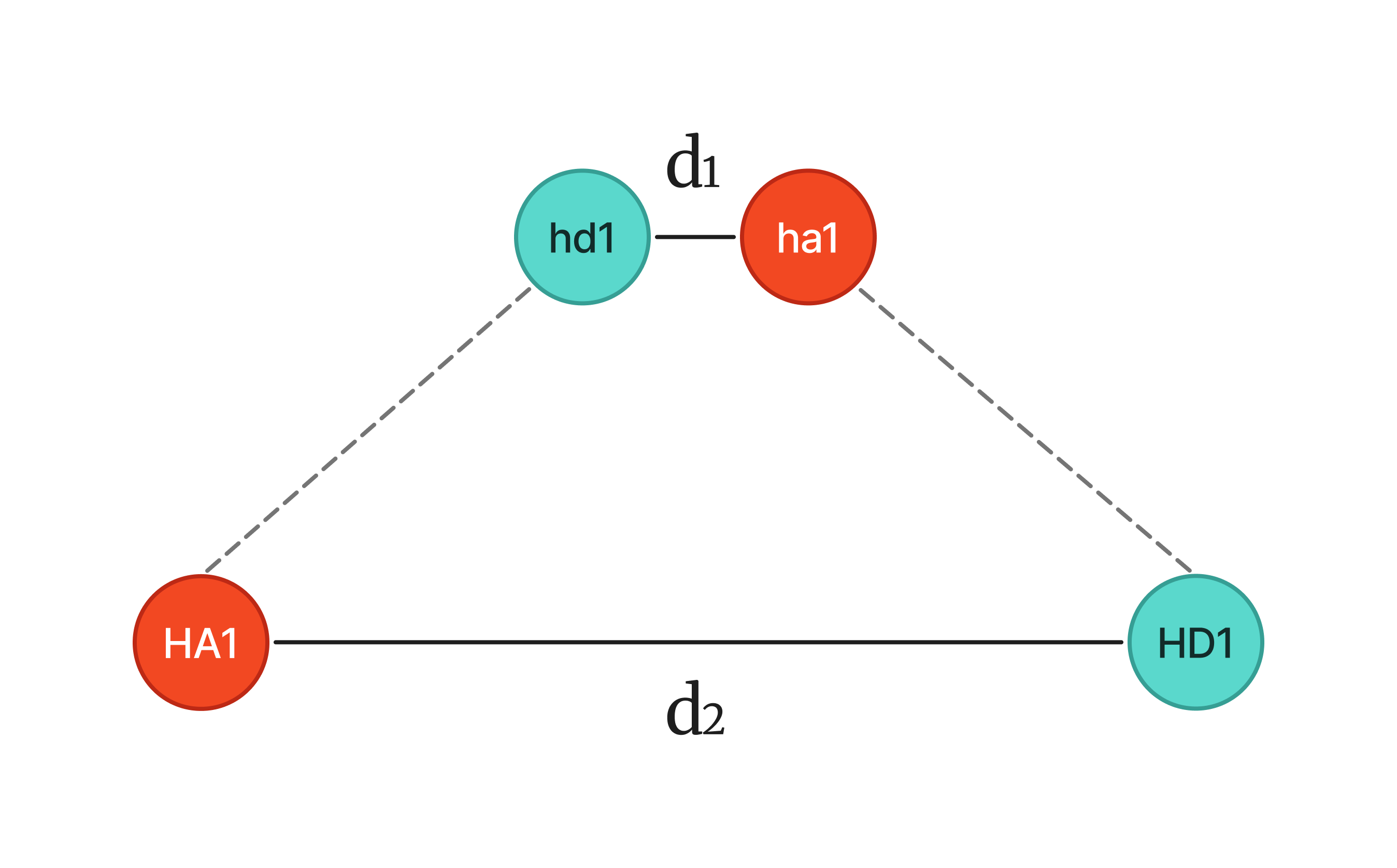}}
  \caption{(a) perfect match, (b) $\tau$-flexible match ($|d_1 - d_2| \geq \tau + \epsilon_1 + \epsilon_2$), and (c
  ) mismatch ($|d_1 - d_2| \leq \tau + \epsilon_1 + \epsilon_2$).}
  \label{fig:match-comparison}
\end{figure}

Figures \ref{fig:match} and \ref{fig:flexible-match} show possible matches, while Figure \ref{fig:mismatch} shows a mismatch.

\begin{figure}[H]
  \centering
  \includegraphics[width=0.6\linewidth]{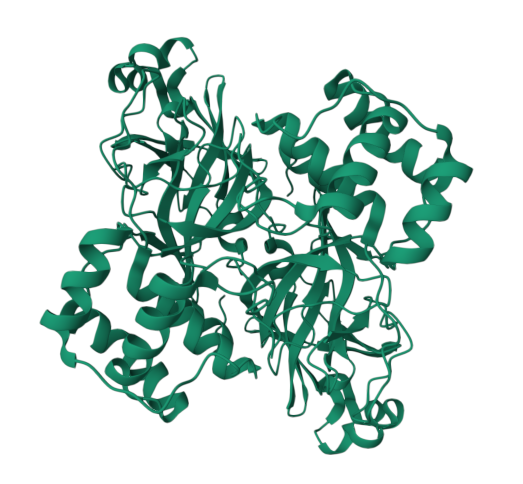}
  \caption{SARS-CoV-2 Mpro complex. Image extracted from PDBbind dataset, PDB ID: 8SKH.}
  \label{fig:SARS-CoV-2 Mpro complex}
\end{figure}

\begin{figure}[H]
  \centering
  \includegraphics[width=0.6\linewidth]{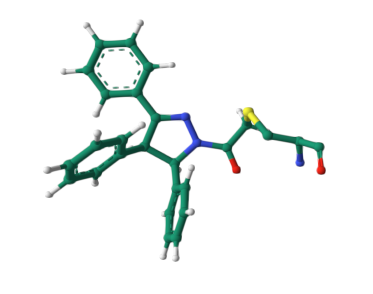}
  \caption{Ligand PM-2-020B. Image extracted from PDBbind dataset, PDB ID: 8SKH.}
  \label{fig:PM-2-020B}
\end{figure}

\begin{figure}[H]
  \centering
  \includegraphics[width=0.6\linewidth]{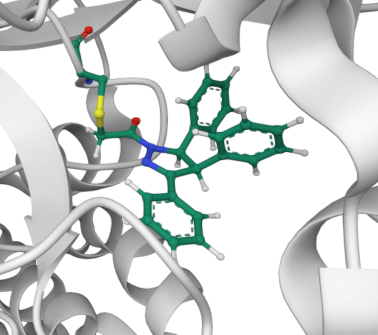}
  \caption{Binding interaction between the receptor (SARS-CoV-2 Mpro complex), shown in color, and the ligand (PM-2-020B), shown in grayscale inset. Image extracted from PDBbind dataset, PDB ID: 8SKH.}
  \label{fig:Protein-ligand interaction}
\end{figure}

Figure \ref{fig:SARS-CoV-2 Mpro complex} shows the SARS-CoV-2 Mpro complex, and Figure \ref{fig:PM-2-020B} shows PM-2-020B. Figure \ref{fig:Protein-ligand interaction} illustrates the interaction between the SARS-CoV-2 Mpro complex and PM-2-020B. All three figures were obtained from PDBbind.

\subsection{Maximum Vertex Weight Clique Formulation}

Under the BIG formulation, the molecular docking problem is transformed into a Maximum Vertex Weight Clique (MVWC) problem, which can be solved through Binary Quadratic Programming (BQP) \cite{PhysRevApplied.21.034036,mvwc-bqp}:

\begin{equation}\label{eq:BQP}
\begin{aligned}
  \text{maximize} \sum_{i = 1}^{N} w_i x_i + \sum_{i = 1}^{N} \sum_{j = 1, j \neq i}^{N} w_{ij} x_i x_j, \\ 
  \text{subject to} \ \  x_i \in \{0,1\} \ \forall i \in \{1,\dots, N\}.
\end{aligned}
\end{equation}

where $N = |V|$ is the number of qubits required to solve the problem, and $x_i$ is a binary variable representing vertex $v_i$. The weight $w_{ij} \ \text{is equal to} \ P \ \text{if} \ \{v_i, v_j\} \in \bar{E}$ and equal to zero otherwise, where $\bar{E}$ is the set of edges of the complement of $\bar{G}$ and P is a negative penalty term.

Using the transformation $x_i = (\sigma_{i}^{z} - 1)/2$ in equation \ref{eq:BQP}, we obtain the QUBO representation:

\begin{equation}
  \mathcal{H} = \frac{1}{2} \sum_{i \in V} w_i(\sigma_{i}^{z}  - 1)+ \frac{P}{4} \sum_{(i, j) \notin E, i\neq j} (\sigma_{i}^{z} - 1)(\sigma_{j}^{z} - 1) 
\end{equation} 

And finally, the solution to the problem is obtained by minimizing $\mathcal{H}$, which is achieved in this manuscript through Quantum Circuit Evolution algorithm, described in the following section.

\subsection{Quantum Circuit Evolution}
\label{QCE}

The main contribution of this manuscript is the use of a quantum evolutionary algorithm for simulating receptor–ligand interactions. Unlike the method based on Gaussian Boson Sampling proposed in the literature \cite{doi:10.1126/sciadv.aax1950}, quantum circuit evolution \cite{Franken2022} can be implemented on gate-based quantum computers. Furthermore, unlike QAOA and DC-QAOA \cite{DC-QAOA-Solano2022}, also used in the molecular docking problem \cite{PhysRevApplied.21.034036}, the proposed method does not rely on the classical optimization of circuit parameters, which may be advantageous for large-scale interaction simulations.

The proposed method begins with a shallow quantum circuit containing only a single quantum gate acting on an arbitrary qubit. Based on this initial circuit, new circuits are generated through the application of the INSERT, DELETE, SWAP, and MODIFY operations, described below, according to predefined probabilities. The initial gate is a single-qubit rotation ($R_x$, $R_y$, or $R_z$) with a random angle $\theta \in [0, 2\pi]$ placed on a randomly chosen qubit.

\begin{itemize}
  \item INSERT: insert a new quantum gate, with a randomly generated parameter $\theta$, at a random position in the circuit;
  \item DELETE: removes a randomly selected gate from the circuit;
  \item SWAP: removes a gate (DELETE) and inserts a new gate (INSERT) at the same randomly selected position in the circuit;
  \item MODIFY: modifies a randomly selected gate in the circuit.
\end{itemize}

For the MODIFY operation, the parameter of the chosen gate is perturbed as $\theta \mapsto \theta + \epsilon$, with $\epsilon \sim \mathcal{N}(0, 0.1)$, so the gate is refined rather than replaced. These four operations act as mutations applied to a single parent circuit, and no recombination between individuals is used.

In this work, a population of 48 individuals was used, such that at each step, 48 circuits are generated through the random application of the operations described above. After generation, the expected value of the objective function in Eq. \ref{eq:Fu} is evaluated, and the circuit with the lowest expected value is selected to generate the next population of circuits:

\begin{equation}\label{eq:Fu}
  \mathcal{F}(U) = \bra{\psi_0} U^\dagger \mathcal{H} U \ket{\psi_0},
\end{equation}
where $\ket{\psi_0}$ is an arbitrary initial state, $U$ is the unitary implemented by the evolved circuit, and $\mathcal{H}$ is the problem Hamiltonian. The objective is the expectation value of $\mathcal{H}$ on the state $U\ket{\psi_0}$, so minimizing $\mathcal{F}(U)$ drives the circuit toward the ground state of $\mathcal{H}$.

Since the best circuit is kept as the parent of the next generation, the selection is elitist and the objective value either decreases or stays the same across generations.

The evolution proceeds until the maximum number of generations or another stopping criterion is reached. In this work, only the maximum number of generations was considered as the stopping criterion.

A complete description of the quantum circuit evolution method is given in our previously work \cite{Fernandez2026}.

\subsection{Simulation Environment}

All experiments were performed using simulations, as the cost of accessing real quantum computers remains high. The simulations were carried out using the CUDA-Q SDK, developed and maintained by NVIDIA. All simulations were executed on the Ogbon HPC cluster at SENAI CIMATEC, Brazil, using only CPU cores from an Intel(R) Xeon(R) Gold 6240 processor.

The only limiting factor for executing these experiments on real quantum computers is the financial cost associated with accessing such devices.

\section{Results}

Table \ref{tab:results} presents the results obtained for the experiments with 6 qubits (8SKH), 8 qubits (3HAC), and 12 qubits (5F4L). All $10$ experiments performed with 6 and 8 qubits converged to the maximum clique. For the 12-qubit case, the optimal solution was found in 4 out of the $10$ experiments, while the remaining $6$ experiments converged to the same clique of size 3, corresponding to a local minimum on the cost function landscape.

\begin{table}[H]
  \centering
  \caption{Experimental results.}
  \label{tab:results}
  \begin{tabular}{|c|c|c|c|c|}
    \hline
    PDB ID & Nº of qubits & Best clique & Clique found \\
    \hline
    8SKH & 6 &  111000 & 111000 \\
    \hline
    3HAC & 8 & 11100001 & 11100001 \\
    \hline
    \multirow{2}{3em}{\ 5F4L} & \multirow{2}{1em}{12} & \multirow{2}{6em}{110000100001} & 110000100001 - 40\%\\ 
    & &  & 000100001100 - 60\% \\
    \hline
    \end{tabular}
\end{table}

\begin{figure}[H]
  \centering
  \includegraphics[width=0.92\linewidth]{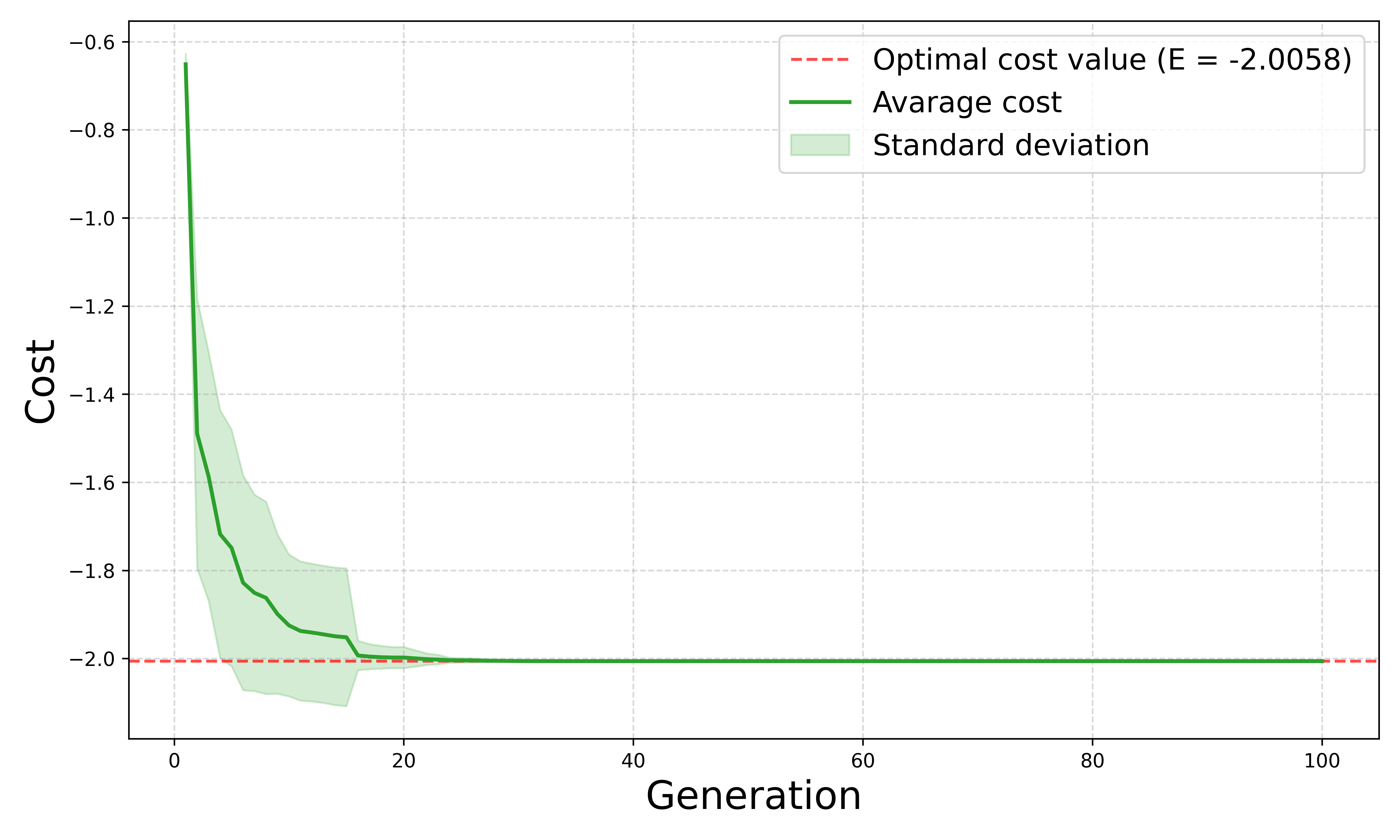}
  \caption{Convergence plot for 8SKH. The red line represents the optimal value of the cost function, computed classically. The green curve represents the mean cost function value across $10$ independent runs, while the shaded region denotes the corresponding standard deviation.}
  \label{Convergence1}
\end{figure}

\begin{figure}[H]
  \centering
  \includegraphics[width=0.92\linewidth]{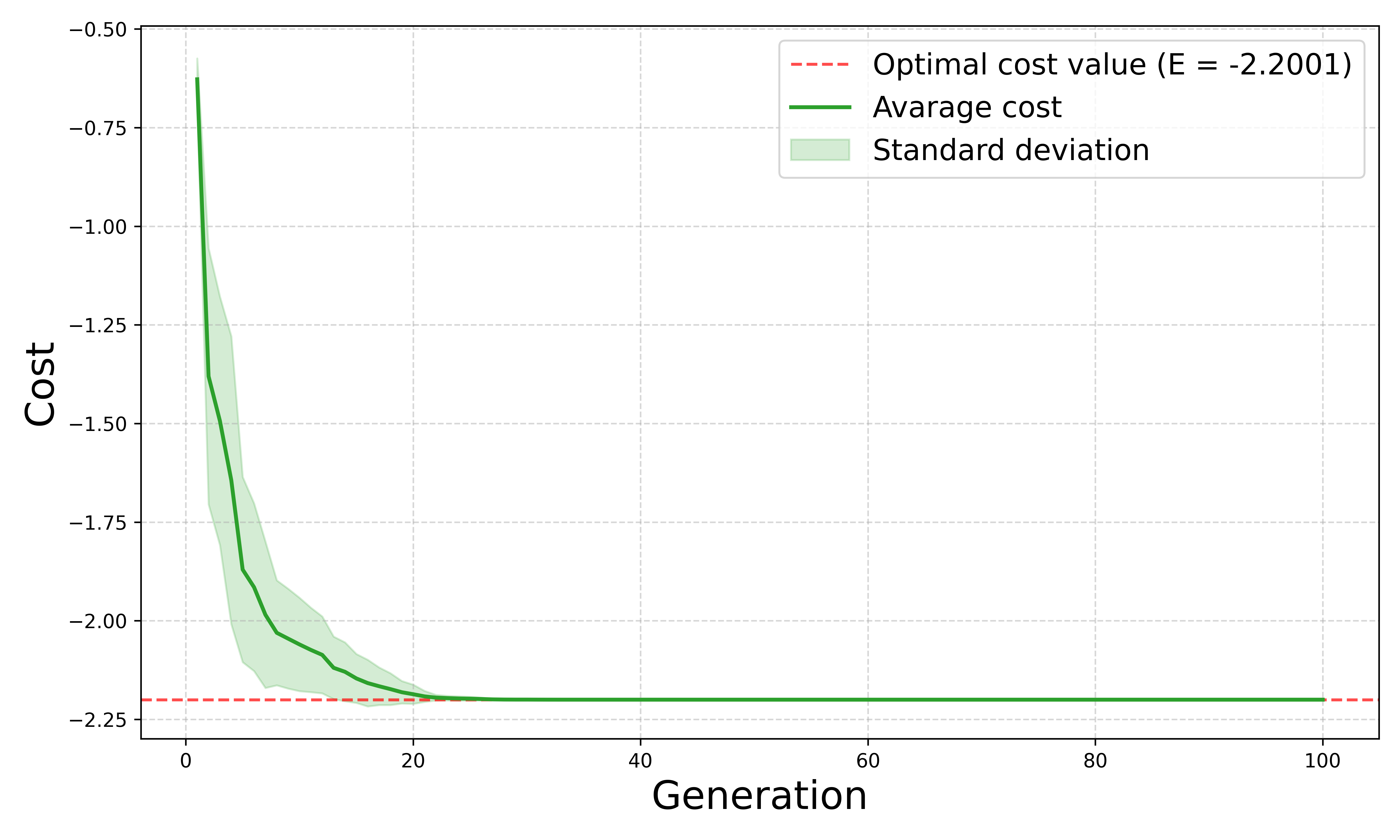}
  \caption{Convergence plot for 3HAC. The red line represents the optimal value of the cost function, computed classically. The green curve represents the mean cost function value across $10$ independent runs, while the shaded region denotes the corresponding standard deviation.}
  \label{Convergence2}
\end{figure}

\begin{figure}[H]
  \centering
  \includegraphics[width=0.92\linewidth]{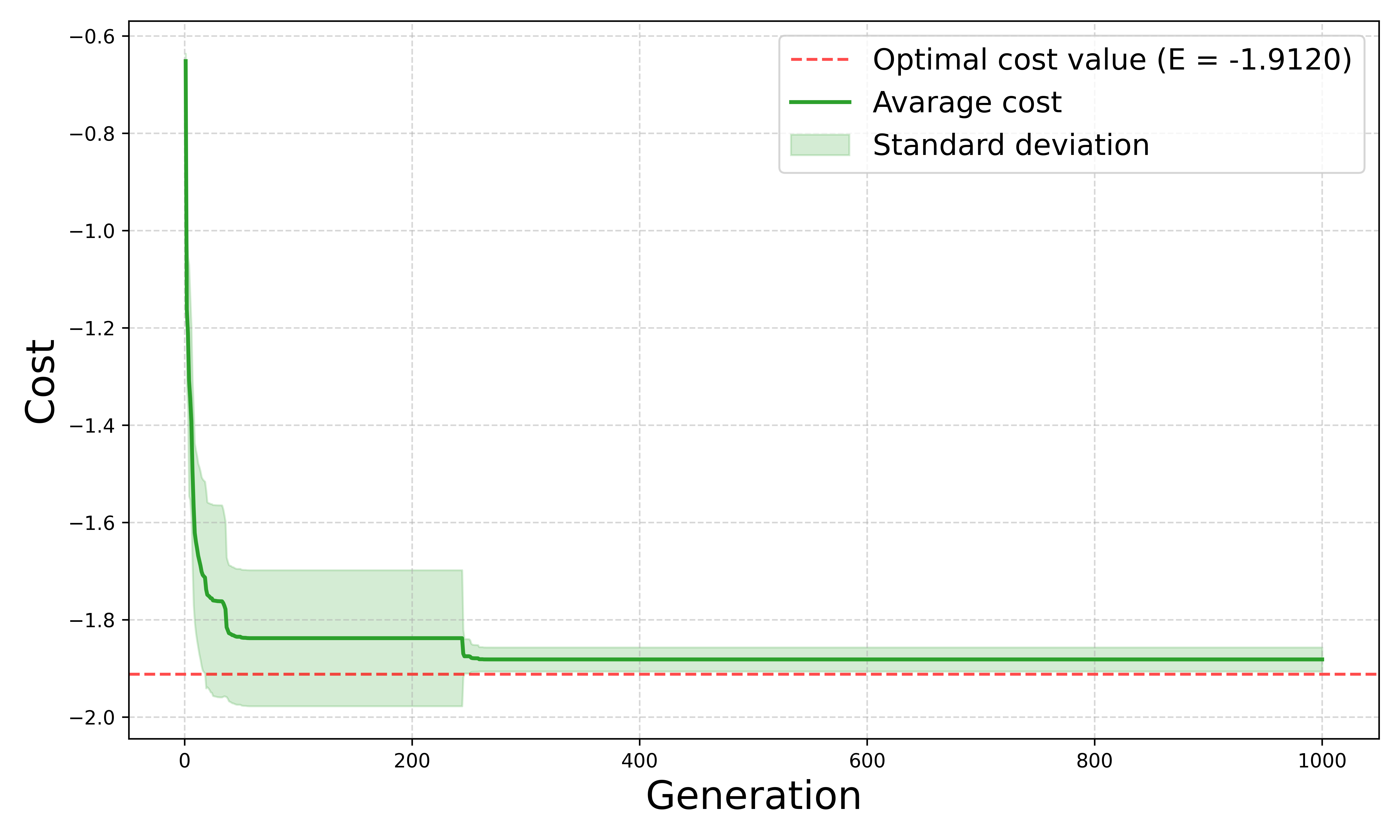}
  \caption{Convergence plot for 5F4L. The red line represents the optimal value of the cost function, computed classically. The green curve represents the mean cost function value across $10$ independent runs, while the shaded region denotes the corresponding standard deviation.}
  \label{Convergence3}
\end{figure}

The convergence behavior of the models is shown in Figures \ref{Convergence1} (8SKH), \ref{Convergence2} (3HAC), and \ref{Convergence3} (5F4L). For the 6 and 8 qubit cases, all experiments converged within fewer than 30 generations. In the 12 qubit case (5F4L), a total of 274 generations was required for all experiments to converge. However, 9 out of the $10$ experiments reached convergence in fewer than 60 generations. The energies for the different cliques can be found in the appendices of Reference \cite{PhysRevApplied.21.034036}.

For comparison, the results reported in \cite{PhysRevApplied.21.034036} exhibit convergence curves that are still decreasing when terminated after 200 iterations for the 6-qubit (8SKH) case and after 1000 iterations for both the 8-qubit (3HAC) and 12-qubit (5F4L) cases. Furthermore, the results presented in \cite{PhysRevApplied.21.034036} indicate that the QAOA found the optimal solution for the 12-qubit (5F4L) model with approximately 50\% of success probability, and DC-QAOA found it with approximately 30\% of success probability. 

Since the same problem formulation was adopted in the present work, these results suggest that the gradient-free evolutionary method presented here exhibits a greater ability to identify the correct solution while requiring a substantially smaller number of iterations.

The circuit depths are presented in Appendix \ref{appendix:supplementary}. Since no additional convergence criterion was adopted, circuit evolution continues until the predefined number of generations has been completed. In all cases, the final circuit depth is greater than the depth of the first generation in which the target solution is found.

In general, the best clique was identified within at most six generations, demonstrating the algorithm's ability to find high-quality solutions using shallow quantum circuits.

\section{Conclusion}
\label{conclusion}

In this work, we present a quantum evolutionary approach to solve the molecular docking problem. Three protein-ligand complexes, previously employed in related studies, were used to enable a direct comparison between the obtained results and those reported in the literature. The results demonstrate that the proposed gradient-free method is capable of consistently finding the optimal solution in only a few optimization steps while eliminating the need for classical parameter optimization. Consequently, the proposed approach overcomes several limitations encountered in previous studies. In future work, we intend to scale the problem to larger protein-ligand complexes and evaluate the method for a higher number of qubits. \\

\textbf{Funding}
This work has been partially supported by the project "Mestrado em tecnologias quânticas" supported by QuIIN - Quantum Industrial Innovation, EMBRAPII CIMATEC Competence Center in Quantum Technologies, with financial resources from the PPI IoT/Manufatura 4.0 of the MCTI grant number 053/2023, signed with EMBRAPII. 
This work has also been partially financed by the "Conselho Nacional de Desenvolvimento Científico e Tecnológico" (CNPq), Brazil, grant number 305096/2022-2 (MAM).

\bibliography{apssamp}

\appendix

\section{Circuit Depths}
\label{appendix:supplementary}

In this section, the circuit depths are presented. For each experiment, the reported values include the circuit depth at the first generation in which the solution is found, the maximum circuit depth reached during the evolutionary process, and the circuit depth at the end of the experiment.

The results of the 10 experiments conducted for 8SKH (Table \ref{tab:depth-8SKH}), 3HAC (Table \ref{tab:depth-3HAC}), and 5F4L (Table \ref{tab:depth-5F4L}) are presented. In the case of 5F4L, Table \ref{tab:depth-5F4L} also reports the clique found, since the obtained solution is not always the global minimum of the cost function.

\begin{widetext}

\begin{table}[H]
  \centering
  \caption{Circuit Depth - 8SKH.}
  \label{tab:depth-8SKH}
  \begin{tabular}{|c|c|c|c|}
    \hline
    Run & Depth first correct & Maximal depth & Final depth (100 steps) \\
    \hline
    1 & 2 & 10 & 4 \\
    \hline
    2 & 2 & 10 & 6 \\
    \hline
    3 & 3 & 10 & 9 \\
    \hline
    4 & 2 & 6 & 5 \\   
    \hline 
    5 & 2 & 7 & 4 \\
    \hline
    6 & 6 & 16 & 7 \\
    \hline
    7 & 2 & 7 & 3 \\
    \hline
    8 & 2 & 9 & 9 \\
    \hline
    9 & 4 & 10 & 9 \\
    \hline
    10 & 3 & 7 & 2 \\
    \hline
    \end{tabular}
\end{table}

\begin{table}[H]
  \centering
  \caption{Circuit Depth - 3HAC.}
  \label{tab:depth-3HAC}
  \begin{tabular}{|c|c|c|c|}
    \hline
    Run & Depth first correct & Maximal depth & Final depth (100 steps) \\
    \hline
    1 & 5 & 10 & 9 \\
    \hline
    2 & 5 & 12 & 10 \\
    \hline
    3 & 3 & 11 & 10 \\
    \hline
    4 & 2 & 8 & 7 \\
    \hline
    5 & 3 & 8 & 4 \\
    \hline
    6 & 5 & 10 & 10 \\
    \hline
    7 & 5 & 11 & 9 \\
    \hline
    8 & 4 & 11 & 7 \\
    \hline
    9 & 4 & 11 & 5 \\
    \hline
    10 & 3 & 8 & 6 \\ 
    \hline
    \end{tabular}
\end{table}

\begin{table}[H]
  \centering
  \caption{Circuit Depth - 5F4L.}
  \label{tab:depth-5F4L}
  \begin{tabular}{|c|c|c|c|c|}
    \hline
    Run & Depth first best & Maximal depth & Final depth (1000 steps) & Clique found \\
    \hline
    1 & 3 & 19 & 14 & 110000100001 \\
    \hline
    2 & 5 & 10 & 6 & 000100001100 \\
    \hline
    3 & 3 & 15 & 7 & 110000100001 \\
    \hline
    4 & 2 & 18 & 7 & 000100001100 \\
    \hline
    5 & 4 & 20 & 10 & 000100001100 \\
    \hline
    6 & 3 & 16 & 4 & 110000100001 \\
    \hline
    7 & 6 & 17 & 11 & 110000100001 \\
    \hline
    8 & 3 & 13 & 4 & 000100001100 \\
    \hline
    9 & 4 & 16 & 12 & 000100001100 \\
    \hline
    10 & 6 & 23 & 8 & 000100001100 \\ 
    \hline
    \end{tabular}
\end{table}

\end{widetext}

\end{document}